\documentclass[aps,prb,twocolumn,showpacs]{revtex4}
\usepackage{latexsym}
\usepackage{amssymb}
\usepackage{graphicx}

\newcommand{\pdag}{{\phantom{\dagger}}}
\newcommand{\bq}{\begin{equation}}
\newcommand{\eq}{\end{equation}}
\newcommand{\bn}{\begin{eqnarray}}
\newcommand{\en}{\end{eqnarray}}

\begin{document}

\title{Kondo-type transport through an interacting quantum dot coupled to ferromagnetic 
leads}
\author{Bing Dong$^{1,2}$, H. L. Cui$^{1}$, S. Y. Liu$^{2}$, and X. L. Lei$^{2}$} 
\affiliation{$^{1}$Department of Physics and Engineering Physics, Stevens Institute of 
Technology, Hoboken, New Jersey 07030 \\
$^{2}$Department of Physics, Shanghai Jiaotong University,
1954 Huashan Road, Shanghai 200030, China}

\begin{abstract}

We investigate the equilibrium and out-of-equilibrium Kondo effects in a single-level 
interacting quantum dot connected to two ferromagnetic leads. Within the non-crossing 
approximation, we calculate the total density of states (DOS), the linear conductance, 
and the nonlinear differential conductance for both the parallel and the anti-parallel 
alignments of the spin polarization orientation in the leads, followed by a brief 
discussion regarding the validity of this approach. Numerical calculations show that 
for the anti-parallel alignment, a single Kondo peak always appears in the equilibrium 
DOS, resulting in the conventional temperature behavior in the linear conductance and 
the zero-bias maximum in the differential conductance. The strength of the DOS peak is 
gradually suppressed with increasing polarization, due to the fact that formation of 
the Kondo-correlated state is more difficult in the presence of higher polarization. On 
the contrary, for the parallel configuration the Kondo peak in the DOS descends 
precipitately and splits into two peaks to form a very steep valley between them. This 
splitting contributes to the appearance of a ``hump" in the temperature-dependent 
linear conductance and a nonzero-bias maximum in the differential conductance. 
Moreover, application of a bias voltage can split each Kondo peak into two in the 
nonequilibrium DOS for both configurations. Finally we point out that the tunnel 
magnetoresistance could be an effective tool to demonstrate the different Kondo effects 
in different spin configurations found here.         

\end{abstract}

\maketitle

\section{Introduction}

The discovery of the Kondo effect in a quantum dot (QD) connected to two normal 
reservoirs has stimulated much experimental and theoretical interests in this many-body 
phenomenon,\cite{Goldhaber} resulting in several novel findings, such as splitting of 
the Kondo peak under nonequilibrium condition,\cite{Schmid} an unususal enhancement of 
conductance in the cases when an even number of electrons resides in a 
QD,\cite{Sasaki,Wiel} and double peaks of the differential conductance in coupled 
double QDs.\cite{Oosterkamp,Jeong} Recently, an increasing attention has been paid, 
because of its potential application in the megnetoelectronics and quantum 
computer,\cite{Prinz} to the spin-polarized electron tunneling through systems 
consisting of two ferromagnetic (FM) leads sandwiched by a QD, which features a 
significant Kondo effect at low temperatures when it connects to normal leads. 
Generally speaking, the Kondo resonance in the density of states (DOS) at the Fermi 
energy originates from screening of the dot spin due to the exchange coupling with the 
conduction electrons. Therefore it is interesting to observe whether the 
Kondo-correlated state can form when the conduction band is of spin polarization, and 
if so, what is the difference from the conventional patterns.

In a recent paper, Sergueev {\it et al.}\cite{Serg} and Zhang {\it et al.}\cite{Zhang} 
presented a theoretical analysis of the transport characteristics of such a FM/QD/FM 
system, using the ansatz proposed by Ng\cite{Ng} and the standard equation-of-motion 
(EOM) technique for the retarded Green's function. They found that there is always a 
sharp single Kondo resonant peak in nonlinear differential conductivity at zero bias, 
regardless of the polarization orientation of the two leads, parallel (P) or 
anti-parallel (AP) configurations. On the contrary, Martinek {\it et 
al.}\cite{Martinek} reported a markedly different result on similar systems, based also 
on an EOM technique and an additional assumption to replace the bare level of QD in the 
resultant self-energy expression with the one self-consistenly determined. They found 
that for P alignment of the lead magnetizations, the Kondo resonances in the DOS split 
for spin-up and -down electrons, thus the differential conductance exhibits 
nonzero-bias maximum and the linear conductance drops to a low value even the 
polarization being as small as 0.2. Later on, L\"u and Liu\cite{Lu} reported also the 
similar splitting by applying Ng's ansatz. Ma {\it et al.}\cite{Ma} employed the 
finite-$U$ slave-boson mean-field approach\cite{KR,Dong} to investigate the 
spin-polarized transport of this system and found drastically different behaviors 
depending on the polarization alignment of the two leads. In view of the ongoing 
controversy, further analysis on the influence of the spin polarization on the 
Kondo-correlated state in a QD is desirable, preferably based on more advanced schemes. 

In this paper, we employ the non-crossing approximation (NCA) based on the auxiliary 
boson technique to carry out a detailed analysis of the FM/QD/FM Kondo problem. The NCA 
is a diagrammatic technique to sum all the non-crossing diagrams in the leading order 
$|V|^2/N$ ($V$ is the hopping matrix element between the local electron and conduction 
electrons, and $N$ is the number of spin degeneracy of the local level).\cite{Bickers} 
And it is proved to be an accurate approach for the case of $N=2$ (of interest for a 
QD), even under out-of-equalibrium conditions.\cite{Cox,Wingreen,Hettler} For a 
symmetric FM/QD/FM system considered in the present paper, although the individual QD 
is exactly spin degenerate, the spin-polarized leads can lift this degeneracy in the 
case of P alignment through the tunneling induced spin-related self-energies. However, 
if the spin polarization strength $p$ of the two leads is small enough, the 
self-energies are of weak spin-dependence, and as a consequence the NCA should be 
adequate to treat this systems. Actually, we calculate the equilibrium and 
nonequilibrium transmissions for the FM/QD/FM system with the polarization strength as 
low as $p=0.2$ for the P configuration. Numerical results show a complete splitting for 
the equilibrium transmission (DOS) and a nonzero-bias maximum for the voltage-dependent 
differential conductance, which are in qualitative agreement with those of 
Martinek\cite{Martinek} and L\"u\cite{Lu}.

The rest of the paper is organized as follows. In section II we describe the model 
Hamiltonian and give a brief description of applications of the infinite-$U$ NCA to the 
spin-dependent transport in FM/QD/FM systems, as well as a qualitative discussion about 
its validity. In section III numerical computations and discussions are presented, 
including the equilibrium and out-of-equilibrium DOS, the linear conductance, and the 
nonlinear differential conductance, for both the P and the AP polarization 
configurations. Finally, all the results are summarized in Section IV.                      

\section{Model and Formulation}

The system Hamiltonian for a quantum dot with a single spin-degenerate energy level 
$\epsilon_{d}$ ($N=2$) connected to two ferromagnetic leads is written as 
\begin{eqnarray}
H&=&\sum_{\eta, k, \sigma}\epsilon _{\eta k\sigma
}c_{\eta k\sigma }^{\dagger }c_{\eta k\sigma }^{\pdag}+ \epsilon_{d} \sum_{\sigma} 
c_{d \sigma }^{\dagger }c_{d \sigma }^{\pdag} \cr
&\hphantom{=}& +Un_{d \uparrow }n_{d \downarrow }+\sum_{\eta, k, \sigma} (V_{\eta 
\sigma}c_{\eta k\sigma }^{\dagger }c_{d \sigma}^{\pdag} 
+{\rm {H.c.}}),
\label{hamiltonian1}
\end{eqnarray}
where $c_{\eta k \sigma}^{\dagger}$ ($c_{\eta k \sigma }$) and $c_{d \sigma}^{\dagger}$ 
($c_{d \sigma}$) are the creation (annihilation) operators for electrons with momentum 
$k$ and spin $\sigma$ in the lead $\eta$ ($={\rm L,R}$) and for a spin-$\sigma$ 
electron on the QD, respectively. The third term describes the Coulomb interaction 
among electrons on the QD, which is assumed to be infinite ($U\rightarrow\infty$) in 
the present paper, forbidding double occupancy. The fourth term represents the 
tunneling coupling between the QD and the reservoirs via
\bq
\Gamma_{\sigma}^{\eta}(\omega)=2\pi \sum_{k} |V_{\eta \sigma}|^2 \delta 
(\omega-\epsilon_{\eta k \sigma}),\,\,\,\,\,\,\,(\eta={\rm L, R}).
\eq
In the wide band limit, $\Gamma_{\sigma}^{\eta}$ is assumed to be constant. For the 
identical leads and symmetric barriers, of interest in the present investigation, the 
ferromagnetism of the leads can be accounted for by the polarization-dependent 
couplings $\Gamma_{\uparrow}^{L}=\Gamma_{\downarrow}^{R}=(1+p)\Gamma_{0}$, 
$\Gamma_{\downarrow}^{L}=\Gamma_{\uparrow}^{R}=(1-p)\Gamma_{0}$ for the P alignment, 
while $\Gamma_{\uparrow}^{L}=\Gamma_{\downarrow}^{R}=(1+p)\Gamma_{0}$, 
$\Gamma_{\downarrow}^{L}=\Gamma_{\uparrow}^{R}=(1-p)\Gamma_{0}$ for the AP alignment. 
$\Gamma_{0}$, and $p$ ($0\leq p< 1$) describe the tunneling coupling between the QD and 
the nonmagnetic leads, and the polarization strength of the leads. Under this 
approximation, the current $I$ through the QD can be expressed in terms of the total 
transmission ${\cal T} (\omega)$ as\cite{Meir}
\begin{equation}
I=\frac{e}{\hbar} \int d\omega [f_{L}(\omega)-f_{R}(\omega)] {\cal T}(\omega),
\label{current}
\end{equation}
where
\bq 
{\cal T}(\omega)=\sum_{\sigma} \frac{\Gamma_{\sigma}^{L}\Gamma_{\sigma}^{R}} 
{\Gamma_{\sigma}^{L}+\Gamma_{\sigma}^{R}}\rho_{\sigma}(\omega),
\label{tran}
\eq
with $\rho_{\sigma}(\omega)=-\frac{1}{\pi} {\rm Im} G_{\sigma}^{r}(\omega)$ being the 
DOS for spin-$\sigma$ electrons. $G_{\sigma}^{r}(\omega)$ is Fourier transform of the 
retarded Green's function,
\bq
G_{\sigma}^{r}(t)=-i\theta(t) \langle {c_{\sigma}(t),c_{\sigma}^{\dagger}(0)} \rangle.
\eq
The main purpose of this work is to calculate the DOS $\rho_{\sigma}(\omega)$ as a 
function of temperature $T$, bare-level energy $\epsilon_{d}$ and bias voltage $V$ for 
different polarization configurations and strengths $p$, and the associated linear and 
nonlinear conductance.

According to the infinite-$U$ slave-boson approach, the ordinary electron operators on 
the QD can be decomposed into a boson operator $b$ and a pseudo-fermion operator 
$f_{\sigma}$,
\bn
c_{d\sigma}(t)&=&b^{\dagger}(t)f_{\sigma}(t), \cr
c_{d\sigma}^{\dagger}(t)&=&f_{\sigma}^{\dagger}(t)b(t),
\en
with a constraint for the auxiliary operators 
$b^{\dagger}b+\sum_{\sigma}f_{\sigma}^{\dagger} f_{\sigma}^{\pdag}=1$.
In the slave boson representation, the Hamiltonian (\ref{hamiltonian1}) for the 
FM/QD/FM systems becomes
\begin{eqnarray}
H&=&\sum_{\eta, k, \sigma}\epsilon _{\eta k\sigma
}c_{\eta k\sigma }^{\dagger }c_{\eta k\sigma }^{\pdag}+ \epsilon_{d} \sum_{\sigma} 
f_{\sigma }^{\dagger }f_{\sigma }^{\pdag} \cr
&\hphantom{=}& +\sum_{\eta, k, \sigma} (V_{\eta \sigma}c_{\eta k\sigma } ^{\dagger } 
b^{\dagger} f_{\sigma} 
+{\rm {H.c.}}).
\label{hamiltonian2}
\end{eqnarray}
In order to evaluate the DOS $\rho_{\sigma}(\omega)$, Wingreen {\it et 
al.}\cite{Wingreen} generalized the NCA to study the nonequilibrium properties of the 
Anderson model connected with two normal conduction bands, using the Keldysh 
nonequilibrium Green's function formalism. It is well-known that the NCA is a 
self-consistent conserving perturbation expansion for the pseudo-fermion and 
slave-boson self-energies to first order in the effective coupling $J=|V|^2$. At the 
lowest order in perturbation diagrams the boson self-energy involves the bare fermion 
propagator while the fermion self-energy involves the bare boson propagator. By 
replacing these bare propagators with the dressed auxiliary particle propagators in the 
Feynman diagram, one can obtain a set of coupled integral equations, which 
self-consistently determine the self-energies of these auxiliary particles. Solving 
these coupled equations is equivalent to summing up a subset of diagrams to all orders 
in $J$. Furthermore, it can be proved that the NCA includes all diagrams of leading 
orders in $1/N$.\cite{Bickers} Therefore, the NCA is expected to be a quantitative 
approach in the limit of large $N$. For a QD connected with normal leads, $N=2$, it is 
already proved to be satisfactory in qualitatively describing the linear and nonlinear 
Kondo-type transport.\cite{Wingreen,Hettler}

Unfortunately, when the level degeneracy is broken, the NCA could produce spurious 
peaks in the DOS, thus is unreliable for transport investigation. For example, as 
mentioned by Wingreen,\cite{Wingreen} the NCA without vertex corrections produces an 
additional Kondo peak at the chemical potential in a finite magnetic field due to a 
false sefl-interaction of each level, whereas other methods find that the Kondo peak 
splits into two peaks. Similarly, the NCA without vertex corrections seems to be also 
inappropriate for the FM/QD/FM systems, because spin-related tunneling lifts the level 
degeneracy in the QD. However, the present situation is somewhat different from 
magneto-transport. The degeneracy lifting is evident in the presence of a magnetic 
field, while is dependent on the relative polarization orientation of the two leads and 
of course the strength of polarization for the FM/QD/FM systems.

When the polarization orientations of the two FM leads is anti-parallel, the 
self-energies are actually independent of spin and the degeneracy remains as two for 
the case of the identical leads and symmetric barriers. On the other hand, for the P 
configuration spin-related tunneling results in the self-energies to be different for 
spin-up and -down electrons. However, it is natural that the deviation depends on the 
polarization strength $p$. Namely, the NCA could still be reliable for the FM/QD/FM 
systems with small enough $p$. Numerical calculations in the next section show that the 
conventional Kondo peak in DOS indeed splits into two peaks completely at the 
polarization $p=0.2$ for the P configuration. Even though one can observe an additional 
peak located at the chemical potential but an order of magnitude smaller than the two 
real Kondo peaks, these results are in agreement with previous 
predictions.\cite{Martinek} Consequently, this convinces us that the NCA provides an 
appropriate description for the symmetric FM/QD/FM systems in either the AP alignment 
with arbitrary polarizations or the P configuration with weak polarizations. Of course, 
this scheme's validity being dependent on the polarization $p$ should be carefully 
checked, for example, by the modified NCA including vertex corrections. This, however, 
entails numerically a much heavier task than the original NCA and is beyond the purpose 
of the present paper. We leave this examination to a future publication.

We outline the formulation employed in this paper as follows. The interested reader can 
refer to Ref.\onlinecite{Wingreen} and Ref.\onlinecite{Hettler} for detail. In the 
slave boson representation, the retarded Green's functions for the boson and 
pseudo-fermions are defined as
\bn
D^{r}(\omega)&=&\frac{1}{\omega-\Pi^{r}(\omega)}, \\
G_{f\sigma}^{r}(\omega)&=&\frac{1}{\omega-\epsilon_{d}-\Sigma_{f\sigma}^{r} (\omega)},
\en
with the corresponding retarded self-energies $\Pi^{r}(\omega)$ and 
$\Sigma_{f\sigma}^{r}(\omega)$. Furthermore, the ``lesser" Green's functions for the 
boson and fermions are related with the ``lesser" self-energies $\Pi^{<}(\omega)$ and 
$\Sigma_{f\sigma}^{<}(\omega)$ as
\bn
D^{<}(\omega)&=&D^{r}(\omega) \Pi^{<}(\omega) D^{a}(\omega),\\
G_{f\sigma}^{<}(\omega)&=&G_{f\sigma}^{<}(\omega) \Sigma_{f\sigma}^{<}(\omega) 
G_{f\sigma}^{a}(\omega).
\en  
The self-consistent NCA equations for out-of-equilibrium are
\bn
\Pi^{r}(\omega)&=&\sum_{\eta={\rm L,R},\sigma} \frac{\Gamma_{\sigma}^{\eta}} {2\pi} 
\int d\varepsilon f(\varepsilon-\omega-\mu_{\eta}) 
G_{f\sigma}^{r}(\varepsilon),\,\,\,\, \\
\Sigma_{f\sigma}^{r}(\omega)&=&\sum_{\eta={\rm L,R}} \frac{\Gamma_{\sigma}^{\eta}} 
{2\pi} \int d\varepsilon f(\varepsilon-\omega+\mu_{\eta}) D^{r}(\varepsilon), \\
\Pi^{<}(\omega)&=& \sum_{\eta={\rm L,R},\sigma} \frac{\Gamma_{\sigma}^{\eta}} {2\pi} 
\int d\varepsilon f(\varepsilon-\omega+\mu_{\eta}) 
G_{f\sigma}^{<}(\varepsilon),\,\,\,\, \\
\Sigma_{f\sigma}^{<}(\omega)&=&\sum_{\eta={\rm L,R}} \frac{\Gamma_{\sigma}^{\eta}} 
{2\pi} \int d\varepsilon f(\varepsilon-\omega-\mu_{\eta}) D^{<}(\varepsilon),  
\en                              
where $f(x)=[\exp(\beta x)+1]^{-1}$ ($\beta=1/k_{\rm B}T$) is the Fermi distribution 
function and $\mu_{\eta}$ is the chemical potential of the $\eta$ lead. After solving 
this set of self-consistent equations, the imaginary part of the retarded local Green's 
function, the DOS $\rho_{\sigma}(\omega)$, can be calculated within the NCA as
\bn
\rho_{\sigma}(\omega)&=&\frac{1}{Z} \int \frac{d\varepsilon}{2\pi} \, 
[D^{<}(\varepsilon) {\rm Im}G_{f\sigma}^{r}(\varepsilon+ \omega) \cr
& + & G_{f\sigma}^{<}(\varepsilon) {\rm Im}D^{r}(\varepsilon - \omega)],
\en
where
\bq
Z=\int d\varepsilon \, [D^{<}(\varepsilon) + \sum_{\sigma} 
G_{f\sigma}^{<}(\varepsilon)].
\eq
Finally we can use Eqs.(\ref{current}) and (\ref{tran}) to calculate the current 
through the QD.    

\section{Numerical results and discussions}

\subsection{Density of State}

In this section we present numerical calculations and discussions. First, we deal with 
the total equilibrium and out-of-equilibrium DOS (transmission) Eq.(\ref{tran}) for the 
FM/QD/FM systems with a fixed bare-level energy $\epsilon_d=-4.0$ ($\Gamma_0$ is used 
as the energy unit throughout the rest of the paper). It is worth noting that the 
systems considered here belong to the deep Kondo regime and are appropriate to 
demonstrate the strong correlated effects.

Fig.\,1(a) shows the total DOS in the AP configuration for several different 
polarizations $p=0,\, 0.2$, and $0.4$ as well as various temperatures $T=0.01,\, 0.02$, 
and $0.04$. Clearly, a significant Kondo peak remains at the Fermi energy (which is 
chosen to be the the energy zero) under the addition of the spin-polarized leads. The 
contribution of spin-polarized leads is to suppress both the Kondo peak and the single 
particle excitation peak (see overall shapes of the DOS in the inset of Fig.\,1). This 
suppression is more pronounced with increasing polarization $p$. These results are 
understandable with the aid of the following considerations. Suppose that in the 
extreme case of two completely spin-polarized leads $p=1$, spin-down electrons are 
completely absent in the left lead, but electrons in the right lead are all spin-up and  
could provide compensation to screen the dot spin and to guarantee the formation of the 
Kondo-correlated singlet state. Thus the Kondo peak still exists with a reduced 
amplitude. Naturally, the cases of weak polarizations $p<1$ are more likely to form the 
Kondo state. Note that increasing temperature can broaden the peak and suppress the 
Kondo resonance as usual. In addition, effects of the external bias voltage on the 
out-of-equilibrium DOS are plotted in Fig.\,1(b). For convenience, we choose a 
symmetric voltage drop such that the chemical potential $\mu_L=-\mu_R=eV/2$ for the 
left and right leads. As expected, we find a splitting of the Kondo peak with a width 
nearly equal to the bias voltage applied between the source and drain leads.            

In short, the QD connected with two AP magnetized leads develops the same Kondo 
resonance as the QD with two normal leads, whereas the former is suppressed to some 
extent depending on the polarization $p$. In contrast, the situation is drastically 
different for the P configuration as shown in Fig.\,2, where we plot the total DOS for 
the polarization $p=0.2$ with various temperatures (a) and bias voltages (b). It is 
clear that the P polarization significantly changes the DOS of the QD in comparison 
with the case of $p=0$. The Kondo resonance splits into two distinct peaks with 
different amplitudes. One moves from the original location $\omega=0$, the Fermi 
energies of the two leads, to a lower energy position, the other shifts to the opposite 
direction at the expense of its height. Moreover the magnitudes of both Kondo 
resonances are largely suppressed by the introduction of spin polarization. We observe 
that there appears a remnant peak located at the Fermi energy, which is produced by the 
NCA calculation without vertex corrections as mentioned above. But its amplitude is one 
order smaller than the two shifted Kondo peaks and can be neglected. As a result, a 
very steep valley is found between the two peaks with a nearly vanished bottom. Finally 
we find that increasing temperature can not only smooth and broaden the peaks as usual, 
but also raise the bottom of the deep valley gradually, which can result in a peculiar 
temperature dependence of the linear conductance as shown in Fig.\,3 (in the next 
subsection).

Fig.\,2(b) depicts effects of changing bias voltage on the total DOS (transmission 
probability) for the same system as in Fig.\,2(a). If we keep the temperature low, 
$T=0.01$, and increase the bias voltage, the two resonances first experience 
suppression and then each of them splits into two distinct peaks. Each pair of the 
peaks has a width about equal to the bias voltage. Increasing the temperature would 
eventually wash out the peak splitting and recover a single but much less pronounced 
peak in both equilibrium and out-of-equilibrium cases.

\subsection{Linear and Nonlinear Conductance}

In Fig.\,3 we plot the calculated linear response conductance $G$ vs. $\log T$ for 
various polarizations $p=0, 0.1, 0.15$, and $0.2$. In the AP configuration (thin line 
in Fig.\,3) the linear conductance $G_{AP}$ exhibits the similar overall temperature 
dependence with those of nonmagnetic leads $p=0$, though smaller in magnitude resulted 
from the suppression of the Kondo resonance as addressed in Fig.\,1(a). For the 
opposite orientation, the conductance $G_{P}$ depends strongly on the polarization 
strength $p$ and exhibits no universal $T$ behavior at low temperatures. Increasing 
$p$, $G_{P}$ is largely suppressed first and develops a ``hump" as a function of 
temperature. This peak is due to the fact that the Kondo resonance is shifted away from 
the Fermi energy as shown in Fig.\,2(a). To demonstrate the dramatic change of 
conductance under different polarization orientations, we plot the tunnel 
magnetoresistance (TMR) defined as $TMR=(G_{P}-G_{AP})/G_{AP}$ in the inset of Fig.\,3. 
We find the TMR arrives at a value as large as $100\%$ at the lowest temperature 
calculated in this work and falls rapidly with increasing $T$, which can be attributed 
to the peculiar Kondo resonance in the P configuration. At high temperatures $G_{P}$ 
approaches the value of nonpolarization $p=0$, leading to a saturated and small 
positive linear TMR.

Fig.\,4 shows the linear conductance as a function of bare-level energy $\epsilon_{d}$, 
which can be tuned via the external gate voltage, for nonpolarization $p=0$, as well as 
the P and the AP configurations with $p=0.2$. As expected, $G_{AP}$ shows the same 
trend with that of nonpolarization though a smaller amplitude. The peak of $G_{p}$, 
however, shifts towards the Fermi energy. This means that the linear TMR changes its 
sign at a certain level energy and has approximatively a symmetric shape around this 
point. Near the empty orbital regime $\epsilon_{d}\simeq 0$ and near the deep Kondo 
regime $\epsilon_{d}\simeq -4$, the linear TMR reaches its maximum value as large as 
$30\%$ in the temperature $T=0.1$.                       

Nonlinear differential conductance $dI/dV$ is believed to be a very useful and 
sensitive tool in experiments to detect the formation of the Kondo-correlated state due 
to its proportionality to ${\cal T}(eV)$ derived from the current formula 
Eq.(\ref{current}), assuming that the total transmission ${\cal T}(\omega)$ 
(nonequilibrium DOS) is unchanged under the external bias voltage $V$. So we illustrate 
in Fig.\,5 the calculated $dI/dV$ under the P (a) and the AP (b) configurations, as 
well as the nonlinear TMR (c) at various temperatures. As pointed out above, because 
electrons with spin-up and spin-down are equally available in the AP configuration, the 
formation of the Kondo-correlated state should not be affected. As a result, all the 
curves in Fig.\,5(b) exhibit a single zero-bias peak and rapid decrease in peak height 
with increasing temperature. When the magnetization is rotated to the P alignment, 
adding bias voltage can greatly enhance $dI/dV$ at low temperatures. This is apparently 
due to the complete splitting of the Kondo peak in the DOS shown in Fig.\,2. The 
nonzero-bias maximum in $dI/dV$ is in good agreement with previous EOM 
calculations,\cite{Martinek} except for the reduced width of splitting and the fine 
shape in the differential conductance. These inconsistences can be attributed to the 
significant change of the nonequilibrium DOS with increasing bias voltage [see 
Fig.\,2(b)]. Furthermore, the nonlinear TMR displays a deep dip in the linear regime 
and changes its sign at a certain bias voltage. This large change in the TMR vs. bias 
voltage reflects the different behaviors of the Kondo resonance in the P and the AP 
configurations.      

\section{Summary}

We have investigated the low-temperature, nonequilibrium properties of a spin-valve 
system consisting of a QD connected to two ferromagnetic leads in the Kondo regime. 
Based on the NCA approach we find markedly different behaviors in the equilibrium DOS 
when changing the relative orientation of spin polarization. In the AP configuration, 
we find that a single Kondo peak always appears through the whole range of polarization 
$0\leq p \leq 1$, as just in a QD connected to two normal leads. Increasing 
polarization $p$ can slightly suppress the amplitude of the peak. In the P 
configuration, the Kondo peak descends greatly and splits completely into two peaks 
even for a weak polarization as low as $p=0.2$, leading to a steep valley with nearly a 
zero bottom. In both configurations the chemical-potential difference (the bias 
voltage) appears in the DOS via the the splitting of the Kondo peak into two peaks. 
Thus four peaks can be found in a moderate bias voltage for the P alignment. Of course 
the amplitudes of these peaks are suppressed by increasing temperature.  

Experimentally, we predict, based on the NCA investigation, that the different Kondo 
effects can be observed in transport through a QD by either linear or nonlinear 
measurements. For the AP configuration, the calculations exhibit the usual temperature 
dependence of the linear conductance and a zero-bias maximum in the nonlinear 
conductance, which are the conventional properties of the Kondo-dominated transport 
through a QD. For the P configuration, however, we find a ``hump" in the 
temperature-dependent linear conductance and a nonzero-bias maximum in the differential 
conductance. These peculiar behaviors are associated with the fact that the Kondo peak 
of the QD is split and shifted away from the Fermi level in the case of the P 
alignment. Furthermore, we suggest that the TMR is a more effective tool to explore the 
different features of the Kondo resonance in different configurations.     
        
Finally, we point out again the applicability of the NCA approach to the FM/QD/FM 
systems. No spin splitting in the case of the AP alignment guarantees that the NCA is a 
reliable approximation for quantities involving the DOS. For the P configuration our 
numerical results for the DOS show satisfactory agreement with previous EOM predictions 
in the case of weak polarization $p=0.2$. Thus we believe that the self-consistent 
second-order perturbation approach provides some qualitative features of the Kondo 
effect in the DOS as long as the polarizations of the leads are weak enough, which can 
serve to furnish a deeper understanding of transport properties in FM/QD/FM systems.     
Advanced NCA studies containing the vertex corrections are required to examine the 
validity of this approach. Work along this line is in progress.

\begin{acknowledgments}
 
B. Dong and H. L. Cui are supported by the DURINT Program administered by the US Army 
Research Office. S. Y. Liu and X. L. Lei are supported by the National Science 
Foundation of China, the Special Funds for Major State Basic Research Project, and the 
Shanghai Municipal Commission of Science and Technology.

\end{acknowledgments}

\begin{figure}[htb]
\includegraphics [width=5cm,height=9.0cm,angle=0,clip=on] {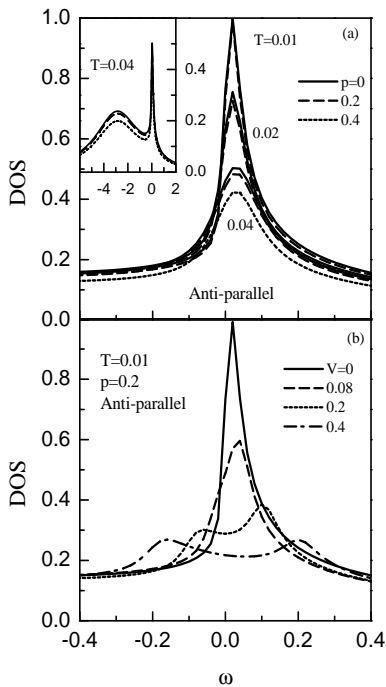}
\caption{The total equilibrium DOS ${\cal T}(\omega)$ in the AP configuration for 
different temperatures $T/\Gamma_{0}=0.01, \,0.02$, and $0.04$ and different 
polarizations $p=0,\,0.2$ and $0.4$; and (b) the nonequilibrium DOS for $T=0.01$ and 
$p=0.2$. The QD in the FM/QD/FM system has a single bare-level energy $\epsilon_d=-4.0$ 
and an infinite on-site Coulomb interaction $U\rightarrow \infty$.} \label{fig1}
\end{figure}


\begin{figure}[htb]
\includegraphics [width=5cm,height=9.0cm,angle=0,clip=on] {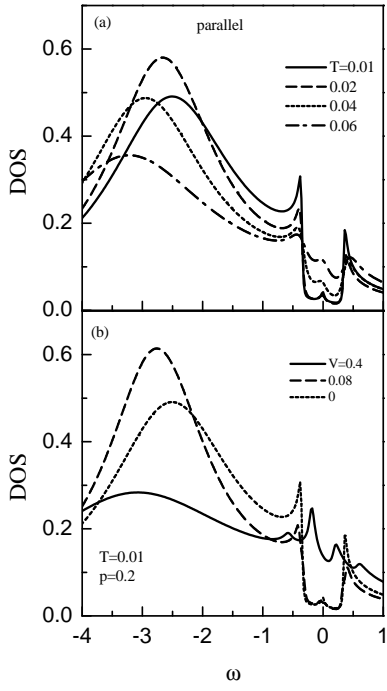}
\caption{(a) The total equilibrium DOS in the P configuration for different 
temperatures $T/\Gamma_{0}=0.01, \,0.02,\,0.04$, and $0.06$; and (b) the nonequilibrium 
DOS for $T=0.01$ and $p=0.2$. The system is the same as described in Fig.\,1.} 
\label{fig2}
\end{figure}


\begin{figure}[htb]
\includegraphics [width=5cm,height=5.0cm,angle=0,clip=on] {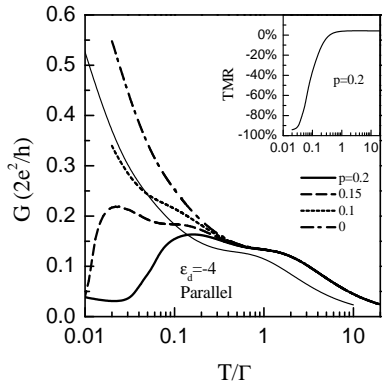}
\caption{The linear conductance $G$ versus temperature at different polarizations $p$ 
for the same system as described in Fig.\,1. The thick lines correspond to the results 
for the P alignment, while the thin curve for the AP alignment at $p=0.2$. Inset: The 
calculated TMR vs. temperature for $p=0.2$} \label{fig3}
\end{figure}


\begin{figure}[htb]
\includegraphics [width=5.cm,height=5.cm,angle=0,clip=on] {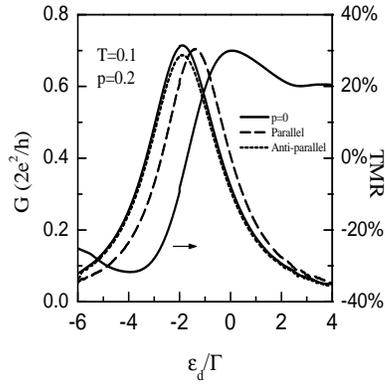}
\caption{The linear conductance $G$ and the TMR versus the bare-level energy of the QD 
at $T=0.1$.} \label{fig4}
\end{figure}


\begin{figure}[htb]
\includegraphics [width=4cm,height=11cm,angle=0,clip=on] {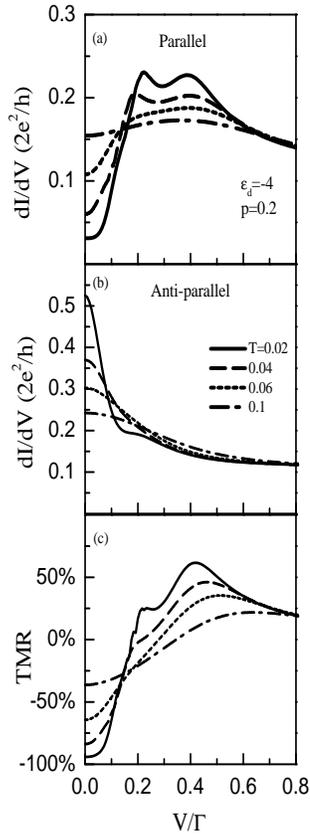}
\caption{The differential conductance $dI/dV$ versus the bias voltage $eV$ at different 
temperatures $T=0.02, \,0.04, \,0.06$, and $0.1$ in the P (a) and the AP (b) 
configurations. (c) The nonlinear TMR versus the bias voltage.} \label{fig5}
\end{figure}

\end{document}